\documentclass[letterpaper, 10 pt, conference]{ieeeconf}
\usepackage{graphicx} 

\IEEEoverridecommandlockouts
\usepackage{graphicx}

\usepackage{physics}
\usepackage{amsmath, amsfonts}
\usepackage{xcolor, soul}
\usepackage[hidelinks]{hyperref}
\usepackage{xurl}

\usepackage{multirow}

\usepackage{algorithmic, algorithm}

\usepackage{textcomp}

\DeclareMathAlphabet{\mathpzc}{OT1}{pzc}{m}{it}

\newcommand{\doa}{\mathcal{X}}
\newcommand{\lyap}{\mathcal{V}}
\newcommand{\idx}[2]{\mathbb{I}_{#1}^{#2}}
\newcommand{\changes}[1]{{#1}}

\DeclareMathOperator*{\argmax}{arg\,max}
\DeclareMathOperator*{\argmin}{arg\,min}

\newcommand\mydots{\hbox to 1em{.\hss.\hss.}}

\title{\LARGE \bf Computationally Efficient Sampling-Based Algorithm\\ for Stability Analysis of Nonlinear Systems
}
\author{Péter Antal$^{1}$, Tamás Péni$^{1}$, and Roland Tóth$^{1,2}$
\thanks{This research was supported by the European Union within the framework of the National Laboratory for Autonomous Systems (RRF-2.3.1-21-2022-00002).}
\thanks{$^{1}$The authors are with the Systems and Control Lab, HUN-REN Institute for Computer Science and Control, Budapest, Hungary (email: antalpeter@sztaki.hun-ren.hu, peni@sztaki.hun-ren.hu, r.toth@tue.nl).}
\thanks{$^{2}$Roland T\'{o}th is also affiliated with the Control Systems Group of the Eindhoven University of Technology, The Netherlands.}}

\begin{document}

\maketitle

\maketitle
\thispagestyle{empty}
\pagestyle{empty}

\begin{abstract}
For complex nonlinear systems, it is challenging to design algorithms that are fast, scalable, and give an accurate approximation of the stability region. This paper proposes a sampling-based approach to address these challenges. By extending the parametrization of quadratic Lyapunov functions with the system dynamics and formulating an $\ell_1$ optimization to maximize the invariant set over a grid of the state space, we arrive at a computationally efficient algorithm that estimates the domain of attraction (DOA) of nonlinear systems accurately by using only linear programming. The scalability of the Lyapunov function synthesis is further improved by combining the algorithm with ADMM-based parallelization. To resolve the inherent approximative nature of grid-based techniques, a small-scale nonlinear optimization is proposed. The performance of the algorithm is evaluated and compared to state-of-the-art solutions on several numerical examples. 
\end{abstract}

\section{Introduction}

Determining the stability region of general nonlinear dynamic systems is one of the most important research directions in the field of control theory \cite{sastry_nonlinear_1999}. To investigate stability properties, Lyapunov analysis is commonly applied, whereby a positive definite function is sought that shows convergence of the system trajectories to a given equilibrium point \cite{lyapunov2011}. If a valid Lyapunov function is found, its level sets can be used to estimate the domain of attraction (DOA) of the equilibrium point, which is a stable region of the system.

Finding a Lyapunov function for general nonlinear systems is difficult, because it mostly involves non-convex optimization. Throughout the past decades, several algorithms have been developed to overcome this issue by either focusing on special system classes (e.g. rational, polynomial), or computing an approximate solution of the optimization problem. 

For example, Sum of Squares (SoS) is a common approach, where the Lyapunov function is constructed as the sum of squared polynomials, inherently ensuring positive definiteness \cite{Parillo2000}. Despite the fact that SoS can be efficiently used for polynomial systems with 2-3 dimensional state space, the optimization becomes computationally intractable for large number of states and general nonlinear systems, limiting the applicability of the algorithm.

To estimate the DOA of uncertain rational nonlinear systems, linear matrix inequality (LMI) conditions have been introduced in \cite{trofino_lmi_2014}. Using the notion of annihilators together with Finsler's Lemma, convex optimization problems are formulated to establish stability properties. Although it is a promising research direction, current results are restricted to rational systems, moreover, they are still either conservative, or computationally overwhelming \cite{trofino_lmi_2014,Polcz2021}.

Recently, sampling-based Lyapunov analysis methods have gained significant attention \cite{ahmed_automated_2020, bobiti_sampling_2016, fernandes_combining_2023}. In \cite{ravanbakhsh_learning_2019}, control Lyapunov functions (CLFs) are synthesized for continuous-time nonlinear systems. The proposed approach consists of a \textit{demonstrator} (generating control inputs from states), a \textit{learner} (proposing candidate Lyapunov functions based on the samples), and a \textit{verifier} (checking if the candidate is valid over the investigated domain of the state space). To verify the proposed candidate, the algorithm is restricted to polynomial dynamics with polynomial CLFs and a semidefinite programming-based relaxation is applied. A similar approach is presented in \cite{chen_learning_2021} to find the DOA of discrete-time nonlinear systems. \changes{This method exploits a neural network approximation of the dynamics and extends quadratic Lyapunov functions by incorporating future states to enrich the quadratic function class. 
Although the algorithm can be applied to general nonlinear systems, it involves the computation of deterministic bounds on the neural network approximation error, which is not yet solved for high dimensional systems.}

\changes{In continuous time, including the system dynamics in the parametrization of the Lyapunov function is known as Krasovskii's method that has been originally used to establish global asymptotic stability \cite{Khalil2002}. In \cite{ahmadi_non-monotonic_2008, ahmadi_non-monotonic_2008_b}, it is shown that the conservativeness of quadratic functions can be reduced by using non-monotonic Lyapunov functions, where the dynamics also appear in the parametrization.}

Learning stability certificates with artificial neural networks is also an emerging field of robotics and control \cite{dawson_safe_2023}. Neural Lyapunov functions have shown remarkable representation capability and they can be trained efficiently \cite{gaby_lyapunov-net_2022,richards_lyapunov_2018}. In \cite{grune_examples_2023}, it is demonstrated that the scalability of Lyapunov function synthesis for certain high dimensional systems can be improved by separable control Lyapunov functions and their neural network approximation. However, due to the difficulties of solving constrained neural network optimizations and verifying neural Lyapunov functions on a continuous domain, their applications are currently limited.


In this paper, we propose a novel iterative, sampling-based approach to estimate the DOA of general continuous-time nonlinear systems. To overcome the limitations of existing methods, our contributions are as follows:
\begin{enumerate}
    \item By turning the Lyapunov function construction as a linear programming problem with $\ell_1$ regularization, we arrive to a computationally efficient method with direct maximization of the invariant set.
    \item We improve the scalability of the Lyapunov function synthesis using the Alternating Direction Method of Multipliers (ADMM) algorithm.
    \item The performance of the proposed method is analyzed and compared to state-of-the-art solutions on several numerical examples.
\end{enumerate}

This work is organized as follows. In Section~\ref{sec:prob}, the problem formulation is introduced. The main contributions of the paper are presented in Sections~\ref{sec:alg} and \ref{sec:admm}, in terms of the algorithm for Lyapunov function synthesis and the ADMM extension for improved scalability. Numerical examples are provided in Section~\ref{sec:examples}, and Section~\ref{sec:conclusion} concludes the paper.

\section{Problem formulation}\label{sec:prob}

We consider nonlinear systems of the form
\begin{equation}\label{eq:sys}
    \dot x(t) = f(x(t)),
\end{equation}
where $x(t) \in \mathbb{R}^n$ is the state vector and $f: \mathbb{R}^n \rightarrow \mathbb{R}^n$ is a 
nonlinear function. A point $x^* \in \mathbb{R}^n$ is an equilibrium point of system \eqref{eq:sys}, if \changes{$f(x^*) = 0$}. The domain of attraction (DOA) associated with an equilibrium point $x^*$ is the set of all initial states, starting from which the trajectories of \eqref{eq:sys} converge to $x^*$. A linear coordinate transformation can shift any equilibrium point to the origin, therefore, without loss of generality (w.l.o.g.), we assume that system \eqref{eq:sys} has an equilibrium point at the origin.

The equilibrium point $x\!=\!0$ of system \eqref{eq:sys} is asymptotically stable, if for each $\epsilon > 0$ and $t_0\in\mathbb{R}$, there exists a $\delta > 0$, such that for all $\| x(0) \| \leq \delta$, $\| x(t) \| \leq \epsilon\ \forall t\geq t_0$ and $\lim_{t\rightarrow \infty} x(t) = 0$, where $\| \cdot\|$ denotes the Euclidean norm. 


Let $\doa$ be a closed subset of $\mathbb{R}^n$ that contains the origin in its interior, moreover, let $\underline \beta,\beta, \bar\beta : [0, b) \rightarrow [0, \infty)$ be strictly increasing functions with $\underline \beta(0) =\beta(0)= \bar\beta(0)=0$ and $b\in\mathbb{R}^+$ such that $\|x\|\leq b, \forall x\in \doa$. If there exists a continuously differentiable function $\lyap : \doa \rightarrow \mathbb{R}$ such that
\begin{subequations}\label{eq:lyap}
    \begin{align}
        & \underline \beta(\|x\|) \leq \lyap(x) \leq \bar\beta(\|x\|) && \forall x\in \doa,\label{eq:lyap:pos}\\
        & \dot\lyap(x) \leq -\beta(\|x\|) &&\forall x\in \doa,\label{eq:lyap:der}
\end{align}
\end{subequations}
then $\lyap$ is called a Lyapunov function (LF), and system \eqref{eq:sys} is asymptotically stable on $\doa$ \cite{isidori1999}. Moreover, the \emph{sublevel sets} of the Lyapunov function given by $\doa_l = \{ x \ | \ \lyap(x) \leq l \}$, $l \in \mathbb{R}$ give an invariant subset of the DOA if $\doa_l \subseteq \doa$.

Throughout this paper, our goal is to find a Lyapunov function such that the maximal level set contained in $\doa$ minimizes the volume of the set difference with respect to the true DOA of the underlying nonlinear system. 

\section{Sampling-based DOA estimation with $\ell_1$ optimization}\label{sec:alg}

In this section, we introduce a sampling-based algorithm for estimating the DOA of system \eqref{eq:sys}. First, we discuss the parametrization of the Lyapunov function. Then, following the concept of counterexample-guided methods (\cite{ravanbakhsh_learning_2019,chen_learning_2021}), we propose an iterative algorithm consisting of a learner and a verifier to construct a Lyapunov function with maximal sublevel set.

\subsection{Lyapunov function parametrization}

We generalize the results of \cite{chen_learning_2021} to continuous-time systems by incorporating the dynamics and its derivatives into the Lyapunov function. \changes{By assuming that $f$ is $d$ times continuously differentiable, moreover, both $f$ and its derivatives are bounded on $\mathcal{X}$, we propose the following parametrization}:
\begin{align}\label{eq:lyap_param}
\begin{split}
        &\lyap(x, P) = z^\top P z, \quad P \in \mathbb{S}^{p},\\ &z = \begin{bmatrix}
        x^\top & f(x)^\top & \dot f(x)^\top & \dots & f^{(d-1)}(x)^\top
    \end{bmatrix}^\top,
\end{split}
\end{align}
where $p=n(d+1)$, $f^{(0)}\!(x)\! =\! f(x)$, and $f^{(i+1)}\!(x)\! =\! \dv t (f^{(i)}(x)) = \frac{\partial}{\partial x}(f^{(i)}(x)) f(x)$ for $i \in \idx{0}{d-2}$. The notation $\idx{i_1}{i_2}$ stands for the set of integers $i_1, i_1 + 1, \dots, i_2 $, and $\mathbb{S}^p$ denotes the set of $p\times p$ symmetric matrices. The derivative of the Lyapunov function given by \eqref{eq:lyap_param} is expressed as follows: 
\begin{align}\label{eq:lyap_param:der}
\begin{split}
        &\dot \lyap(x, P) = \dot z^\top P z + z^\top P \dot z,\\
        &\dot z = \begin{bmatrix}
        f(x)^\top & \dot f(x)^\top & \ddot f(x)^\top & \dots & f^{(d)}(x)^\top
    \end{bmatrix}^\top.
\end{split}
\end{align}
From \eqref{eq:lyap_param}-\eqref{eq:lyap_param:der}, it follows that both $\lyap(x)$ and $\dot\lyap(x)$ are linear in $P$ when the state is fixed, which we exploit later to efficiently synthesize such Lyapunov functions. \changes{Moreover, boundedness of $z, \dot{z}$ ensures that the r.h.s. of \eqref{eq:lyap:pos} is automatically satisfied.}

Choosing the value of $d$ generates a tradeoff between the complexity of the Lyapunov function synthesis and the accuracy of the DOA estimation. In practice, it is recommended to start with $d=1$ and increase its value until good coverage of the DOA is obtained.



\subsection{Learner based Lyapunov synthesis}

The goal of the learner is to propose a candidate Lyapunov function. The Lyapunov conditions in their original form given by \eqref{eq:lyap} are infinite dimensional, therefore the problem is too complex to solve directly. Instead, we employ a grid-based approach, and synthesize a candidate Lyapunov function that satisfies \eqref{eq:lyap} at discrete points of the state space. Later, we show how the inherent approximative nature of the resulting grid-based problem can be overcome.

As the first step, we define $\mathbb{X}$ as the region of interest (ROI) of the system, in which we would like to locate the DOA. \changes{In general, we assume that $\mathbb{X}$ is available based on prior knowledge about the specific system.} In this paper, we consider $\mathbb{X}$ to be a hyperrectangle, formulated as follows:
\begin{equation}\label{eq:grid}
    \mathbb{X} = \{ x \in \mathbb{R}^n \ | \ \underline x^{(i)} \leq x^{(i)} \leq \bar x^{(i)}, \ i=1\dots n\}
\end{equation}
where $\underline x, \bar x \in \mathbb R^n$ are lower and upper bounds on the state variables. Next, we generate a data set $\mathcal{D}_N$ by drawing $N$ samples on a grid over $\mathbb{X}$ as initial conditions and simulate the system dynamics until convergence to a small neighborhood of the origin or divergence is reached. Based on the simulations, we partition $\mathcal{D}_N$ into subsets $\mathcal{X}_0$ and $\mathcal{X}_\infty$, $\mathcal{D}_N = \doa_0 \cup \doa_\infty$, where $\mathcal{X}_0$ contains the stable initial conditions, $|\mathcal{X}_0| = N_0$, and $\mathcal{X}_\infty$ contains the unstable initial conditions, $|\mathcal{X}_\infty| = N - N_0 = N_\infty$. The notation $| \cdot |$ stands for the cardinality of a given set.

The learner proposes a Lyapunov function candidate and a DOA estimate in terms of the largest sublevel set contained in $\doa$ considering the sample points in $\mathcal{D}_N$.  As the value $l$ associated with the largest sublevel set $\doa_l= \{ x \ | \ \lyap(x,P) \leq l \}$ can be tuned arbitrarily with the scaling of $P$, w.l.o.g. we choose $l=1$. The best solution would be to separate $\doa_0$ and $\doa_\infty$ by the contour line $\lyap(x,P)=1$, i.e., $\lyap(x, P) \leq 1\ \ \forall x \in \mathcal{X}_0$ and $\lyap(x, P) > 1\ \ \forall x \in \mathcal{X}_\infty$. However, if the a priori fixed structure of $\lyap$ does not allow this perfect separation, we still intend to exclude all elements of $\doa_\infty$ from the level set, while trying to include from $\doa_0$ as many points as possible. To solve this problem, we apply $\ell_1$ optimization, as follows:


\begin{subequations}\label{eq:lasso_lf}
    \begin{align}
        P^* =& \argmin_{P\in\mathbb{S}^{p}, \alpha\in\mathbb{R}^{N_0}} \|\alpha\|_1 = &&\!\!\!\!\!\!\!\!\argmin_{P\in\mathbb{S}^{p}, \alpha\in\mathbb{R}^{N_0}}\sum_{i=1}^{N_0} \alpha_i\\
        \text{s.t.} \quad & \alpha_i \geq 0,   && \forall i\in \idx{1}{N_0},\label{eq:lasso_lf:1}\\
        & \lyap(x_i, P) \leq 1+\alpha_i,  && \forall x_i \in \mathcal{X}_0, i\in \idx{1}{N_0}, \label{eq:lasso_lf:con_pos__}\\
        & \lyap(x_i, P) \geq \varepsilon x_i^\top x_i,  && \forall x_i \in \mathcal{X}_0, i\in \idx{1}{N_0}, \label{eq:lasso_lf:con_pos_}\\
        & \dot\lyap(x_i, P) \leq - \varepsilon x_i^\top x_i, && \forall x_i \in \mathcal{X}_0, i\in \idx{1}{N_0}, \label{eq:lasso_lf:con_der}\\
        & \lyap(x_j, P) \geq 1+\delta, \quad && \forall x_j \in \mathcal{X}_\infty, j\in \idx{1}{N_\infty}.\label{eq:lasso_lf:con_pos}
    \end{align}
\end{subequations}
By minimizing the $\ell_1$ norm, we enforce a sparse solution, i.e., maximize the number of zero elements of $\alpha \in \mathbb{R}^{N_0}$ \cite{boyd_convex_2004}. Due to Constraint~\eqref{eq:lasso_lf:con_pos__}, each zero entry corresponds to a stable point contained in the level set. Furthermore, Constraints~\eqref{eq:lasso_lf:con_pos_}, \eqref{eq:lasso_lf:con_der} represent the Lyapunov conditions given by \eqref{eq:lyap}, where $\varepsilon \in \mathbb{R}^+$ is chosen to be a small constant\changes{\footnote{The quadratic function class for $\underline \beta,\beta$ is a standard choice, but other parametrizations could also be applied as an extension of Optimization \eqref{eq:lasso_lf}.}}. Moreover, Constraint~\eqref{eq:lasso_lf:con_pos} ensures that the unstable points are outside of the 1 level set by a margin of $\delta \in \mathbb{R}^+$. Generally, $\delta \approx 10^{-2}$ is a good choice, but selecting a higher value (e.g. 0.1-0.2) often results in improved numerical stability. \changes{In fact, handling $\delta$ as a decision variable could improve feasibility in exchange for increased complexity of the optimization.}

However, $\dot\lyap(x_i, P)$ is forced to be negative even at points which are outside of the sublevel set, which can lead to infeasibility of the optimization. To solve this issue, we modify Constraint~\eqref{eq:lasso_lf:con_der} to prescribe negativity only within the 1 level set of $\lyap(x, P)$, i.e., where $\alpha_i=0$, as follows:
\begin{align}
\dot\lyap(x_i, P) \leq \alpha_i - \varepsilon x_i^\top x_i, \quad \forall x_i \in \mathcal{X}_0, i\in \idx{1}{N_0}.\label{eq:lasso_lf_2:roi_der}\tag{6e\textquotesingle}
\end{align}
Optimization \eqref{eq:lasso_lf} is linear in $P$ and $\alpha$, therefore it is a \emph{linear program} (LP), which can be solved efficiently by numerical solvers. If the problem is feasible, the result gives a Lyapunov function candidate that is valid on the sample set.

The number of optimization variables in \eqref{eq:lasso_lf} is $n_\mathrm{opt}=n^2(d+1)^2 + N_0$. Assuming $n_\mathrm{g}$ number of grid points in each dimension, $N_0$ grows polynomially with $n_\mathrm{g}$ and exponentially with $n$. Hence, scalability is mostly limited by the number of state variables and grid points, while it is less affected by the value of $d$.

\subsection{Verifier optimization}

The candidate Lyapunov function proposed by the learner satisfies 
\eqref{eq:lyap} only at the sample points, therefore a verifier is required additionally to make sure that the Lyapunov conditions 
are satisfied between the samples, as well. For this, the following two optimization problems are solved:
    \begin{align}\label{eq:verifier}
        \gamma^* = \max_{x \in \bar{\doa}} \dot\lyap(x, P^*),\quad \eta^* = \min_{x \in \bar{\doa}} \lyap(x, P^*),
    \end{align}
where $\bar{\doa} = \doa_1 \setminus \{0\}$, $\doa_1$ is the sublevel set $\lyap(x, P) \leq 1$, and $P^*$ is the solution of $\eqref{eq:lasso_lf}$. Optimization \eqref{eq:verifier} is nonlinear as the system dynamics appear in the objective function. However, the number of optimization variables is always $n$ regardless of $d$ and $N$, therefore it can be solved efficiently, e.g. by using CasADi 
\cite{Andersson2019}. Considering the result of Optimization~\eqref{eq:verifier}, if $\gamma^*\! <\! 0$ and $\eta^*\! >\! 0$, we accept $\lyap(x, P^*)$ as a valid Lyapunov function. Otherwise, we add the corresponding $x$ value (the \textit{counterexample}) to $\doa_0$ and call the learner to compute a new candidate. The iteration between the learner and the verifier terminates either if a valid LF is found, or a specified maximum number of iterations ($i_\mathrm{max}$) is reached. The main steps of the proposed sampling-based DOA estimation method are outlined in Algorithm~\ref{alg:iter}.

\begin{figure}[t]
\begin{algorithm}[H]
\caption{Sampling-based iterative DOA estimation}
\label{alg:iter}
\begin{algorithmic}[1]
    \STATE \textbf{input:} $\doa_0, \doa_\infty$
    \STATE Let $i=0$
    \WHILE{$i<i_\mathrm{max}$}
    \STATE \textbf{solve} \eqref{eq:lasso_lf} to obtain $P^*$
    \STATE \textbf{solve} \eqref{eq:verifier} to obtain $\gamma^*, \eta^*$
    \IF {$\gamma^* \geq 0$ }
    \STATE $\doa_0 \leftarrow \doa_0 \cup \argmax_{x \in \bar{\doa}} \dot\lyap(x, P^*)$
    \ELSIF {$\eta^* \leq 0$}
    \STATE $\doa_0 \leftarrow \doa_0 \cup \argmin_{x \in \bar{\doa}} \lyap(x, P^*)$
    \ELSE
    \RETURN $P^*$
    \ENDIF
    \STATE $ i \leftarrow i+1$
	\ENDWHILE
 \RETURN NaN
\end{algorithmic}
\end{algorithm}
\end{figure}

\section{Improving scalability}\label{sec:admm}

The scalability of the learner optimization is important to address, because assuming constant grid resolution, the number of samples grows exponentially with the dimension of the state space. Specifically, Constraints~\eqref{eq:lasso_lf:con_pos__}, \eqref{eq:lasso_lf_2:roi_der} can be written in the form of $A \xi \leq b$ with $\xi = [\ \mathrm{vec}(P)^\top\ \alpha^\top\ ]^\top \in \mathbb{R}^{p^2 + N_0}$, $A \in \mathbb{R}^{2N_0 \times (p^2 + N_0)}$, and $b \in \mathbb{R}^{2N_0}$. For a 5 dimensional system with $d=1$ and 10 samples along each dimension, this means that $N_0=10^5$, therefore $A$ has more than $10^{10}$ elements. The solution of such a large scale problem is very difficult without distributing the constraints.

To overcome the scalability issues, we employ the Alternating Direction Method of Multipliers (ADMM) \cite{boyd_distributed_2010}. For this, Optimization~\eqref{eq:lasso_lf} can be rewritten in the following more compact form:
\begin{align}\label{eq:admm_init}
        \xi^* = &\argmin_{\xi \in C} J(\xi),
\end{align}
where $J$ is the linear cost function, and $C$ is the set of affine constraints. We express $C$ as the intersection of $m$ affine sets, i.e., $C=C_1 \cap C_2 \cap \dots \cap C_m$, such that Constraints~\eqref{eq:lasso_lf:1}, \eqref{eq:lasso_lf:con_pos_}, \eqref{eq:lasso_lf:con_pos} are included in all $m$ sets, and Constraints~\eqref{eq:lasso_lf:con_pos__}, \eqref{eq:lasso_lf_2:roi_der} are divided into $m$ parts.

By introducing $\tilde \xi = [\ \xi_1\ \xi_2\ \mydots\ \xi_m\ ]$, $ \tilde z = [\ z_1\ z_2\ \mydots\ z_m\ ]$, $ \tilde \xi, \tilde z \in \mathbb{R}^{(p^2 + N_0) \times m}$, 
\eqref{eq:admm_init} can be rewritten in the standard form used, e.g., by \cite{boyd_distributed_2010} to derive the ADMM steps, as follows:
\begin{align}\label{eq:admm_ind}
\begin{split}
            \min_{\tilde \xi = \tilde z} \sum_{i=1}^m J(\xi_i) + I_C(\xi_1,\mydots,\xi_m) + I_D(z_1, \mydots, z_m),
\end{split}
\end{align}
where $I_C, I_D$ are indicator functions of the sets $\tilde C = C_1 \times C_2 \times \dots \times C_m$, and $D = \{\tilde z \in \mathbb{R}^{(p^2 + N_0) \times m}\ |\ z_1 = z_2 = \dots = z_m\}$, respectively. Then, using the scaled form of ADMM \cite{boyd_distributed_2010}, the following update rules are formulated:
\begin{subequations}
    \begin{align}
        &\tilde \xi^{k+1} = \argmin_{\tilde{\xi} \in \tilde C} \sum_{i=1}^m J(\xi_i) + \frac{\rho}{2} \| \tilde \xi - \tilde z^k + \tilde u^k \|^2,\label{eq:admm_1}\\  
        &\tilde z^{k+1} = \argmin_{\tilde{z} \in D} \frac{\rho}{2} \| \tilde \xi^{k+1} - \tilde z + \tilde u^k \|^2,\label{eq:admm_2}\\
        &\tilde u^{k+1} = \tilde u^k + \tilde \xi^{k+1} - \tilde z^{k+1},
    \end{align}
\end{subequations}
where $\tilde u$ is the scaled dual variable, and $\rho \in \mathbb{R}^+$ is the step size parameter. 
Optimization~\eqref{eq:admm_1} can be decoupled and parallelized, as follows:
\begin{align}
    \xi_i^{k+1} = \argmin_{\xi_i \in C_i} J(\xi_i) + \frac{\rho}{2} \| \xi_i - \tilde z_i^k + \tilde u_i^k \|_2^2,\ i\in \mathbb{I}_1^m.
\end{align}
The update of $\tilde z$ in \eqref{eq:admm_2} can be solved in closed form:
\begin{align}
    \tilde z^{k+1} = [\ \bar\xi\ \bar\xi\ \dots\ \bar\xi\ ], \quad \bar\xi = \frac{1}{m} \sum_{i=1}^m (\xi_i^{k+1} + u_i^k).
\end{align}
The convergence of the iterative algorithm can be examined by monitoring the following two residual terms:
    \begin{align}
        r^k = \tilde \xi^k - \tilde z^k,\quad s^k = \rho (\tilde z^{k+1} - \tilde z^k).
    \end{align}
If the norm of both residual terms are small for a given $K\in \mathbb{Z}^+$, i.e., $\| r^K \| \leq \bar\varepsilon$ and $\| s^K \| \leq \bar\varepsilon$ for a user-defined $\bar\varepsilon \in \mathbb{R}^+$, then $\xi^K$ is returned as the solution of the original problem given by \eqref{eq:admm_init}.

In practice, selecting the value of $m$ needs to be considered carefully. While choosing a large value leads to good scalability, convergence of the algorithm can become slow as it is also pointed out in \cite{boyd_distributed_2010}. On the other hand, small values of $m$ result in faster convergence, but the scalability improvement is less significant. Generally, the minimal value of $m$ for which the optimization can be solved is a good choice.

\section{Examples}\label{sec:examples}


\begin{table*}
        \caption{Parameters and results of the numerical examples}
\centering
    \begin{tabular}{c|c|c|c|c|c|c|c|c|c|c|c|c|c}
      \multirow{2}{*}{system} & \multirow{2}{*}{$n$} & \multirow{2}{*}{true DOA volume}   & \multirow{2}{*}{$d$} & \multirow{2}{*}{$\varepsilon$} & \multirow{2}{*}{$\delta$} & \multirow{2}{*}{$N$} & \multirow{2}{*}{iterations} & \multicolumn{3}{c|}{estimated DOA volume} &  \multicolumn{3}{c}{computation time} \\
      & & & & & & & & ours & \cite{trofino_lmi_2014} & \cite{Polcz2021} & ours (sec.) & \cite{trofino_lmi_2014} (min.) & \cite{Polcz2021} (sec.)\\
      \hline \hline
       \eqref{eq:vanderpol} & 2 & 66.84 & 2 & $10^{-3}$ & $0.15$ & 900 & 2 & 57.72 & 50.61 & - & 0.093 & 1452 & - \\
      \hline
       \eqref{eq:sys2} & 2 & 8.736 & 1 & $10^{-3}$ & $0.1$ & 900 & 1 & 8.44 & 6.96& - & 0.015 & 522 & - \\
       \hline
        \eqref{eq:sys3} & 2 & 17.28 & 3 & $10^{-3}$ & $0.1$ & 900 & 3 & 16.39 & 15.38 & - & 0.273 & 760 & - \\
        \hline
       \eqref{eq:sys4} & 3 & 888.12 & 1 & $10^{-4}$ & $0.1$ & 9261 & 8 & 529.49 & 32.02 & 314.48 & 2.830 & 561 & 4.758 \\
       \hline
       \eqref{eq:5d} & 5 & 31754.63 & 1 & $10^{-2}$ & $0.5$ & 59049 & 5 & 5087.06 & - & - & 986.67 & - & - \\
       \hline
    \end{tabular}
    \label{tab:my_label}
\end{table*}

We evaluate the performance of the proposed algorithm by estimating the DOA of a number of nonlinear dynamic systems, all of which have an asymptotically stable equilibrium point at the origin. First, we examine two- and three-dimensional systems taken from \cite{trofino_lmi_2014, Polcz2021}. Then, we show how ADMM can be used to construct the Lyapunov function and estimate the DOA of a five-dimensional system. The numerical values of the parameters, computation times, and number of iterations are collected in Table~\ref{tab:my_label}. To help the interpretation of our results, we also indicate the true DOA of each system, obtained by simulating the dynamics on a dense grid. All computations have been performed on a laptop with Intel Core i7 CPU, Ubuntu 20.04 OS, and 32 GB of RAM. To solve the convex learner optimization and the nonlinear verifier optimization, we have used Mosek\footnote{\url{https://mosek.com/}} and CasADi \cite{Andersson2019}, respectively. All of our source code used for the numerical study is available at \url{https://github.com/AIMotionLab-SZTAKI/sampling_based_lyapunov}.

\begin{figure}
    \centering
    \includegraphics[width=\linewidth]{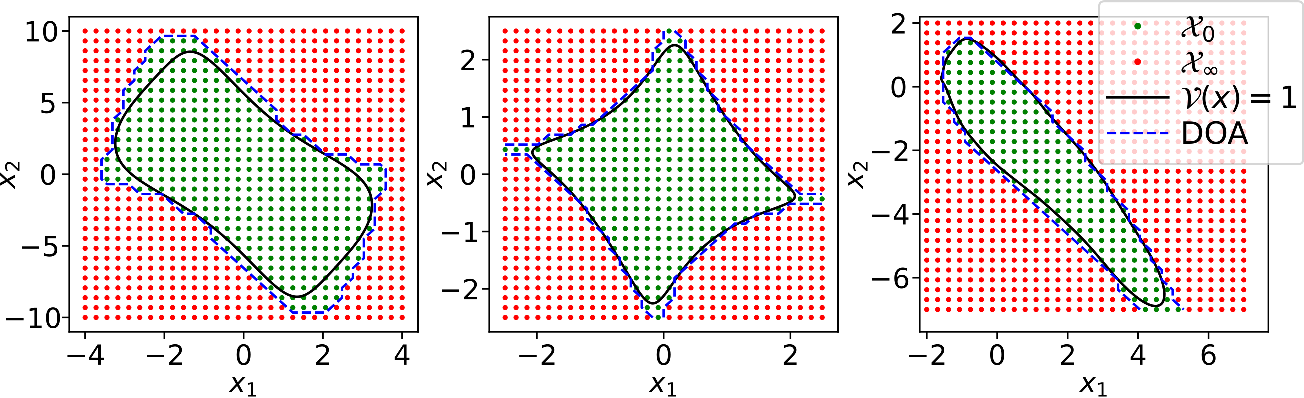}
    \caption{Estimated domain of attraction for multiple 2D systems. Green and red points denote the elements of $\mathcal{X}_0$ and $\mathcal{X}_\infty$, respectively, while the solid black and dashed blue lines correspond to the 1 level set of the Lyapunov function and the true DOA of the system. Left: Van der Pol system given by \eqref{eq:vanderpol}, center: Example 2 given by \eqref{eq:sys2}, right: Example 3 given by \eqref{eq:sys3}.}
    \label{fig:vanderpol}
\end{figure}

\subsection{Two-dimensional examples}

\subsubsection{Van der Pol oscillator}

As the first example, we have examined the DOA of a Van der Pol oscillator, governed by the following dynamics:
\begin{equation}\label{eq:vanderpol}
    \begin{bmatrix}
        \dot x_1 \\ \dot x_2
    \end{bmatrix} = \begin{bmatrix}
        x_2 \\ -2 x_1 - 3 x_2  +x_1^2 x_2
    \end{bmatrix}.
\end{equation}
The ROI has been chosen to $\mathbb{X} = [-4, 4] \times [-10, 10]$, similarly to \cite{trofino_lmi_2014}. Samples have been drawn on a uniform grid over $\mathbb{X}$ using 30 grid points in each dimension, resulting in $N=900$ samples collected to $\mathcal{D}_N$. 

The results are displayed on the left panel of Fig.~\ref{fig:vanderpol}, where the level set of the Lyapunov function and the true DOA are depicted by the solid black and dashed blue lines, respectively, while the green and red points denote the elements of $\mathcal{X}_0$ and $\mathcal{X}_\infty$. After only 2 iterations, the algorithm has returned the following coefficient matrix:

{\footnotesize \begin{align*}
P_1^*\! =\! \begin{bmatrix}
0.1192 & 0.068 & 0.0 & 0.0 & -0.0022 & -0.0001\\
0.068 & 0.0 & 0.0 & 0.0 & 0.004 & 0.0\\
0.0 & 0.0 & 0.0606 & 0.0 & 0.0 & 0.0004\\
0.0 & 0.0 & 0.0 & 0.0 & -0.0006 & 0.0\\
-0.0022 & 0.004 & 0.0 & -0.0006 & 0.0 & 0.0\\
-0.0001 & 0.0 & 0.0004 & 0.0 & 0.0 & 0.0
\end{bmatrix}
\end{align*}}
Fig.~\ref{fig:vanderpol} shows that the DOA estimate is very accurate, few stable points are outside of the sublevel set. Moreover, the numerical results show that our method outperforms \cite{trofino_lmi_2014} both in terms of the volume of the DOA estimation and the solution time required to compute the Lyapunov function.

\subsubsection{Example 2}
The second example is Eq. (143) of \cite{trofino_lmi_2014}:
\begin{equation}\label{eq:sys2}
    \begin{bmatrix}
        \dot x_1 \\ \dot x_2
    \end{bmatrix} = \begin{bmatrix}
        - x_1 +x_1 x_2^2 \\x_1 - x_2 + x_1^2 x_2 - x_1 x_2^2
    \end{bmatrix}.
\end{equation}

In this case, we have generated 900 initial samples on a uniform grid over the domain $\mathbb{X} = [-2.5, 2.5] \times [-2.5, 2.5]$. The algorithm has successfully found a Lyapunov function and DOA estimate in only 1 iteration, resulting in the following coefficient matrix:

{\footnotesize
\begin{align*}
P_2^* = \begin{bmatrix}
0.0 & -0.0021 & 0.0097 & 0.1424\\
-0.0021 & 1.0534 & 0.3903 & 0.4305\\
0.0097 & 0.3903 & 0.0973 & 0.0325\\
0.1424 & 0.4305 & 0.0325 & 0.0294
\end{bmatrix}.
\end{align*}}
The center panel of Fig.~\ref{fig:vanderpol} shows that the DOA of the nonlinear system is estimated accurately in this case, as well. Similar to the first example, the results outperform \cite{trofino_lmi_2014} both in terms of the volume of the DOA estimation and the computation time.


\begin{figure}
    \centering
    \includegraphics[width=.8\linewidth]{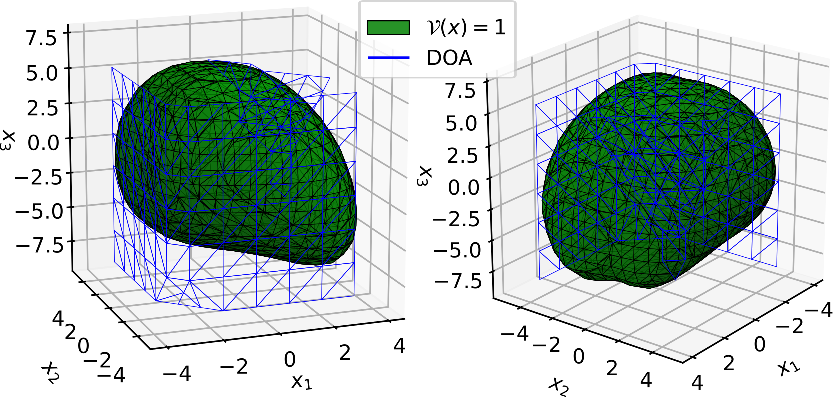}
    \caption{DOA estimation of the 3D example given by \eqref{eq:sys4}. The green surface depicts the level set of the LF, and the blue mesh shows the DOA.}
    \label{fig:3d} 
\end{figure}

\subsubsection{Example 3}
The third example is more challenging than Examples 1 and 2, because it has an asymmetric domain of attraction. The dynamic model of the system is described by Eq. (137) in \cite{trofino_lmi_2014}, as follows:
\begin{align}\label{eq:sys3}
    \begin{bmatrix}
        \dot x_1 \\ \dot x_2
    \end{bmatrix} = \begin{bmatrix}
        x_2 + p_1 \zeta(x) \\- x_1 - x_2 + p_2 x_1^2 
    \end{bmatrix},\quad \zeta(x) = \frac{x_1}{x_2^2 + 1},
\end{align}
where $p_1=p_2=0.5$ are used as numerical values of the parameters. The ROI has been chosen to $\mathbb{X} = [-2, 7] \times [-7, 2]$, and similar to the other 2D examples, 900 initial samples has been generated on a uniform grid (30 along each dimension). For an accurate DOA estimation, 2 derivatives of the dynamics have been included in the Lyapunov function, resulting in 64 parameters of the coefficient matrix. Although the solve time of this example is higher than previous ones due to the increased number of parameters, it is still only 0.273 s, which shows excellent performance of the algorithm. The numerical value of the coefficient matrix obtained after 3 iterations is as follows:

{\tiny \begin{align*}
P_3^*\! =\! \begin{bmatrix}
1.18 & 0.55 & 0.0 & 1.47 & 0.0 & 1.03 & 0.0 & -0.03\\
0.55 & 0.89 & -1.23 & 2.01 & -0.82 & 0.85 & 0.56 & 0.81\\
0.0 & -1.23 & 0.79 & -3.83 & 2.57 & -0.52 & -0.24 & -0.71\\
1.47 & 2.01 & -3.83 & -0.12 & -1.55 & -1.37 & -0.01 & -0.51\\
0.0 & -0.82 & 2.57 & -1.55 & 4.36 & 1.66 & 0.0 & 0.53\\
1.03 & 0.85 & -0.52 & -1.37 & 1.66 & 0.23 & 0.06 & 0.06\\
0.0 & 0.56 & -0.24 & -0.01 & 0.0 & 0.06 & 0.0 & 0.01\\
-0.03 & 0.81 & -0.71 & -0.51 & 0.53 & 0.06 & 0.01 & 0.02
\end{bmatrix}.
\end{align*}}
In the right panel of Fig.~\ref{fig:vanderpol}, it is shown that most of the stable points are contained in the level set of the Lyapunov function, despite the unique asymmetric shape of the DOA.

\subsection{Three-dimensional example}\label{sec:3d}

As the next example, we have chosen a three-dimensional system that is examined both in \cite{trofino_lmi_2014} and \cite{Polcz2021}:
\begin{align}\label{eq:sys4}
    \begin{bmatrix}
        \dot x_1 \\ \dot x_2 \\ \dot x_3
    \end{bmatrix} = \begin{bmatrix}
        x_2 + p_3 x_3 + p_1 \zeta(x) \\- x_1 - x_2 + p_2 x_1^2 \\ p_3 (-2 x_1 - 2 x_3 - x_1^2)
    \end{bmatrix},\ \zeta(x) = \frac{x_1}{x_2^2 + 1},
\end{align}
where the numerical values of the parameters have been set to $p_1=p_2=p_3=0.5$. To estimate the DOA, 21 initial samples have been generated along each dimension of the ROI $\mathbb{X} = [-4, 4] \times [-5, 5] \times [-8.5, 7]$, resulting in 9261 points. The coefficient matrix of the resulting Lyapunov function is

{\footnotesize \begin{align*}
P_4^*\! =\!\begin{bmatrix}
0.063 & -0.015 & 0.0 & -0.056 & 0.003 & 0.007\\
-0.015 & -0.092 & -0.051 & 0.085 & 0.01 & 0.0\\
0.0 & -0.051 & 0.0 & 0.0 & 0.0 & 0.0\\
-0.056 & 0.085 & 0.0 & -0.12 & -0.057 & -0.051\\
0.003 & 0.01 & 0.0 & -0.057 & -0.005 & -0.008\\
0.007 & 0.0 & 0.0 & -0.051 & -0.008 & 0.003
\end{bmatrix}.
\end{align*}}

The computation time is larger compared to the 2D examples, due to the higher number of iterations required to find the Lyapunov function, and the increased dimension of the LP. The 1 level set is a 3 dimensional surface, which is displayed in Fig.~\ref{fig:3d}. Compared to \cite{trofino_lmi_2014}, the volume of our DOA estimate is more than 10 times larger, while the computation time is lower by multiple orders of magnitude. The DOA estimation of \cite{Polcz2021} is significantly better than the one found in \cite{trofino_lmi_2014}, but it is also outperformed by our results both in terms of computation time and volume of the estimated DOA.

\subsection{Five-dimensional example with ADMM}\label{sec:5d}

We analyze the scalability of the algorithm with ADMM extension on a 5 dimensional system, which has been synthesized by coupling the dynamics of \eqref{eq:vanderpol} and \eqref{eq:sys4}, as follows:
\begin{equation}\label{eq:5d}
        \begin{bmatrix}
        \dot x_1 \\ \dot x_2 \\  \dot x_3 \\ \dot x_4 \\ \dot x_5
    \end{bmatrix} = \begin{bmatrix}
        x_2 \\ -2 x_1 - 3 x_2  +x_1^2 x_2 - x_4 \\ x_4 + p_3 x_5 + p_1 \zeta(x) \\- x_3 - x_4 + p_2 x_3^2 \\ p_3 (-2 x_3 - 2 x_5 - x_3^2)
    \end{bmatrix},\quad \zeta(x) = \frac{x_3}{x_4^2 + 1},
\end{equation}
where $p_1=p_2=p_3=0.5$ have been used, similarly to the three-dimensional example of Sec.~\ref{sec:3d}. The two parts of the dynamics are coupled by the $-x_4$ term of the second equation. Consequently, \eqref{eq:5d} can be written in the form of
\begin{subequations}
    \begin{align}
        \dot{\mathpzc{x}}_1 &= f_1 (\mathpzc{x}_1, \mathpzc{x}_2), && \mathpzc{x}_1 = [x_1\ x_2]^\top,\\
         \dot{\mathpzc{x}}_2 &= f_2 (\mathpzc{x}_2), && \mathpzc{x}_2 = [x_3\ x_4\ x_5]^\top.
    \end{align}
\end{subequations}
Given that the origin of both \eqref{eq:vanderpol} and \eqref{eq:sys4} are asymptotically stable, i.e., $f_1(0, \mathpzc{x}_2) = 0$ and $f_2 (0) = 0$, the origin of the resulting coupled system given by \eqref{eq:5d} is also asymptotically stable. For the proof of stability of the coupled system with stable subsystems, see e.g. \cite{isidori1995nonlinear}.

\begin{figure}
    \centering
    \includegraphics[width=\linewidth]{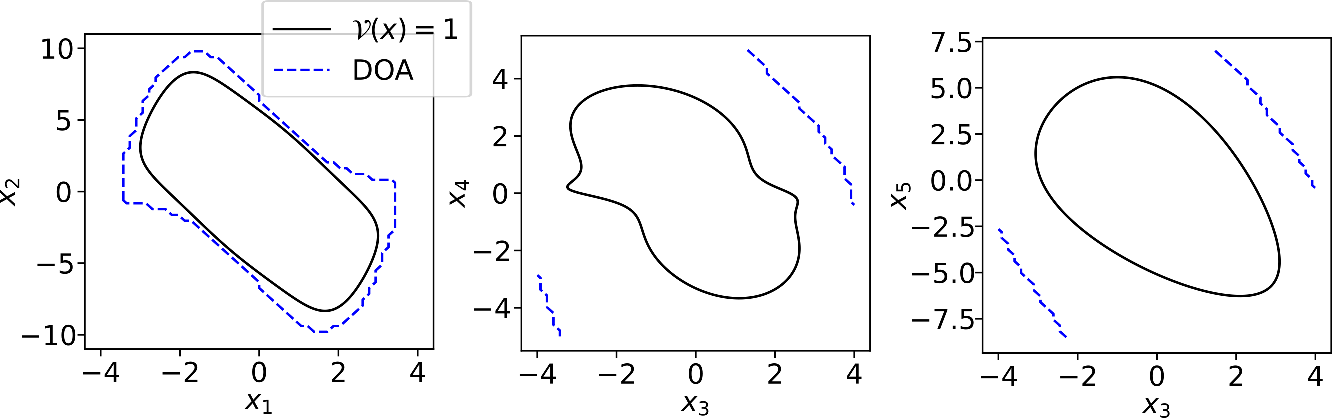}
    \caption{DOA estimation of the 5D example given by \eqref{eq:5d}. Each plot shows a slice of the level set $\lyap(x)=1$ and the DOA, depicted by solid black and dashed blue lines, respectively.}
    \label{fig:5d}
\end{figure}

We have determined $\mathbb{X}$ by combining the ROI of systems \eqref{eq:vanderpol} and \eqref{eq:sys4}, resulting in $\mathbb{X} = [-4, 4] \times [-10, 10] \times  [-4, 4] \times [-5, 5] \times [-8.5, 7]$. Then, we have generated a uniform grid of 9 sample points along each dimension, i.e., 59049 samples altogether to simulate the system dynamics and construct $\doa_0, \doa_\infty$. The learner optimization has been infeasible due to the high number of constraints. To make the optimization tractable, we have employed the ADMM algorithm detailed in Section~\ref{sec:admm}. Using ADMM, we have split the constraints to $m=2$ parts and run the two optimizations in parallel. This way, the problem has been solved successfully.

To construct the Lyapunov function, 5 iterations have been performed between the learner and the verifier, moreover, ADMM required 18 iterations in average to solve the learner optimization, resulting in 16 minutes of solve time altogether. The DOA estimation is illustrated in Fig.~\ref{fig:5d}, where slices of the level set are depicted together with slices of the DOA. The left panel depicts the slice at $x_3=x_4=x_5=0$, showing that the DOA estimation is very accurate. The center and right panels show that the corresponding slice of the true DOA is not contained in the ROI, therefore the curves are not closed. However, the level set of the LF covers a significant part of the DOA in these planes, as well. 

It is important to note that the increased computation time of this example is mainly due to the high number of ADMM iterations. However, on the one hand, most existing DOA estimation methods are limited to 3 or 4 dimensional systems (see e.g. \cite{trofino_lmi_2014,bobiti_sampling_2016,fernandes_combining_2023,chen_learning_2021}), therefore the successful solution already shows improvement compared to these algorithms. On the other hand, the implementation of ADMM could be optimized further to improve the convergence and reduce the overall computation time.

\section{Conclusion}\label{sec:conclusion}

In this paper, we have presented a novel sampling-based algorithm to estimate the domain of attraction of nonlinear systems. By applying $\ell_1$ optimization to maximize the volume of the invariant set of the constructed Lyapunov function, our algorithm have generated accurate DOA estimates quickly and efficiently compared to state-of-the-art techniques. Furthermore, the ADMM extension has improved the scalability of the algorithm, resulting in successful DOA estimation of a five dimensional nonlinear system.

In future research, the scalability of the algorithm could be further improved by large-scale convex optimization methods. This way, our approach could be extended to handle even more complex and high-dimensional nonlinear systems.

\bibliographystyle{IEEEtran}
\bibliography{reference}

\end{document}